\documentstyle[12pt,aaspp4]{article}
\def\<{\langle}
\def\>{\rangle}
\def\avg#1{\< #1 \>}
\def\etal{{et. al. }}
\def\keV{\rm keV}

\begin{document}
\title{The Hardness Distribution of Gamma-Ray Bursts}
\author{ Ehud Cohen, Tsvi Piran }
\affil{Racah Institute of Physics, The Hebrew University, Jerusalem 91904, Israel }
\and
\author{Ramesh Narayan}
\affil{Harvard-Smithsonian Center for Astrophysics, Cambridge, MA 02138, U.S.A.
 }

\begin{abstract}
It is often stated that gamma-ray bursts (GRBs) have typical energies
of several hundreds $\keV$, where the typical energy may be
characterized by the hardness H, the photon energy corresponding to
the peak of $\nu F_{\nu}$.  Among the 54 BATSE bursts analyzed by Band
et al. (1993), and 136 analyzed by us, more then 60\% 
have $50\keV < H < 300\keV$. Is the narrow range of H a real feature
of GRBs or is it due to an observational difficulty to detect harder
bursts?  We consider a population of standard candle bursts with 
a   hardness distribution: $\rho(H)d \log H \ \propto \ H^{\gamma} d
\log H$ and no luminosity - hardness correlation. We model the detection
algorithm of BATSE as a function of H, including cosmological effects,
detector characteristics and triggering procedure, 
and we calculate the expected distribution
of H in the observed sample for various values of $\gamma$.  Both
samples shows a paucity of soft (X-ray) bursts, which may be
real. However, we find that the observed samples are consistent with a
distribution above $H=120\keV$ with $\gamma \sim -0.5$ (a slowly
decreasing numbers of GRBs per decade of hardness). Thus, we suggest
that a large population of unobserved hard gamma-ray bursts may exist.
\end{abstract}

\eject
\section{Introduction}
One striking feature that is common to all gamma-ray bursts (GRBs) is
the fact that most of the observed photons correspond to low energy
gamma-rays, with energies of a few tens to few hundreds of
$\keV$. While other features of the bursts, in particularly the
temporal structure, vary significantly from one burst to another, this
feature seems to be quite invariant.  One wonders, therefore, whether
this is a clue to the nature of GRBs - a phenomenon that theorists
should strive to explain - or if it is just a consequence of an
observational bias against detection of harder or softer bursts.  In
other words, one can ask whether the observed hardness distribution
represents the real one.

\cite{NP95} have  assumed a simple model for the 
sources and the detector, and used a sample of 54 relatively strong
bursts analyzed  by \cite{band}, to find out if GRBs have
intrinsic hardness values around $100-400\keV$, or the observed
distribution is just a data selection effect.  They have found
that the intrinsic hardness distribution can be extended to include
hard bursts with no upper limit.

We calculate the expected observed hardness distribution for several
intrinsic hardness distributions. The calculations include
cosmological red-shift and detector characteristics.  We calculate the
observed hardness distribution of a set of 136 bursts  and we
compare the theoretical distribution to the observed one. We examine
which intrinsic distributions are consistent with the data and which
are not.

In section \ref{obs_data} we describe our data set, the method used
for estimating the spectra and the resulting hardness distribution.
In section \ref{theo_hard} we calculate the expected observed hardness
distribution from a given intrinsic distribution.  As our calculations
deal with cosmological effects on the hardness distribution, we
include in section \ref{corr_dis} a discussion of the possible
correlation between intensity and hardness of cosmological bursts.
Finally, in section \ref{conc} we discuss the constrains imposed on
the intrinsic distribution by the observed data.

\section{The Observed Hardness Distribution}
\label{obs_data}
\subsection{Data}
Using a count spectrum averaged over the estimated total duration
interval for each event, we calculate the  photon energy spectra for a 
group of GRBs using the MER/CONT data from BATSE Large Area
Detectors.  These data consist of count rates in 16 energy channels
spanning a range of approximately $20-2000\keV$, with different
temporal resolutions. To estimate the bursts' spectra we must
subtract the background. This background was fitted with a polynom
of order one or two on intervals before and after the burst
(\cite{Nem95}). 

We limit our sample to bursts that occurred before November 1991 and
between February 1992 and January 1993. We consider bursts with a
minimal peak-flux condition ( $flux_{256}>0.5 \ ph \ cm^{-2} s^{-1}$) and
require the availability of flux measurements in the 256msec, counts
in the 1024msec channel and an estimate of the burst duration.  We
consider only bursts that have continuous data for all the duration of
the burst, (the data dropouts is due to telemetry conditions and has
no relation to the bursts' data, we expect, therefore, that this
sample is a proper sample of the GRB population.).  A total of 136
bursts satisfy these conditions.  
        
\subsection{The Intrinsic Spectrum}
\label{spect_estim}
The BATSEs' LAD detector estimates the energy of incident
photons. However, due to various detector characteristics
(\cite{Pend95}), there is no one to one correspondence between the
true energy of the photon and the measured one.  The BATSE team
provides for each burst a $DRM$ matrix which describes the
detector response to photons at various energies, i.e.
\begin{equation}
C=DRM*P,
\end{equation}
where $P$ is the incident photon spectra ( vector length is 62),
$DRM$ is the detector response matrix ( size 16*62) and 
$C$ is the count spectra (a vector of length 16).

The counts spectra must be transformed into a photon spectra. A direct
inversion is impossible as it is well known that the inverse matrix is
singular.  
We have used the forward folding method. This is a model dependent
method.  One assumes that the photon spectra is well described by a
given functional shape with some unknown parameters, (we have used the
Band parameterization).  For a given set of parameters, the assumed
spectral form is integrated into the $DRM$ spacing, multiplied by the
$DRM$, and compared with the measured count vector. Then, we use the
$\chi^2$ optimization method to find the parameters that fit best the
measured count vector.

\subsection{The Band Spectrum}
With the necessity of a assuming a spectral form, we follow
\cite{band}, by characterizing the bursts' spectra using a four
parameter function:
\begin{equation}
\label{band_spectra}
N_P(E)dE = \cases{ {A / 100\keV } \left [{E / 100\keV} \right ]^{\alpha}
e^{-E/E_0} & $E<(\alpha-\beta)E_0$ \cr {A / 100\keV} \left [{ (\alpha-\beta)E_0
/ 100\keV } \right ]^{\alpha-\beta}e^{\beta-\alpha} \left [ {E / 100\keV} \right ]^\beta &
$E>(\alpha-\beta)E_0$ } 
\end{equation}
This function, which provides a good fit to most of the observed
spectra, is characterized by two power laws joined smoothly at a break
energy $(\alpha - \beta)E_0$. For most of the observed values of
$\alpha$ and $\beta$, $\nu F_{\nu} \ \propto \ E^2N(E)$ rises below
$H=(\alpha+2)E_0$, and decrease above it. The energy $H$ is thus the
``typical'' energy of the observed burst.  Note that the hardness
ratio in BATSE catalogue, which is the ratio of photons observed in
channel 3 to those observed in channel 2, is different from $H$ defined
in this way.

The total energy of a burst described by this spectral form depends on
the hardness of the burst, and on its power-law slopes. 
Using $\gamma(a > 0 ,x) = \int_0^x e^{-t}
t^{a-1} dt$, we calculate the total energy of a burst
\begin{equation}
E_{TOT} = \int_0^{\infty} E N_P(E) dE = A E_0 ({E_0 \over
100\keV})^{\alpha+1} \left [ \gamma( \alpha + 2, \alpha-\beta ) -
(\alpha-\beta)^{\alpha+2} e^{\beta-\alpha}/ (\beta+2) \right ].
\label{total_energy}
\end{equation}

\begin{figure}
\begin{center}
\plotone{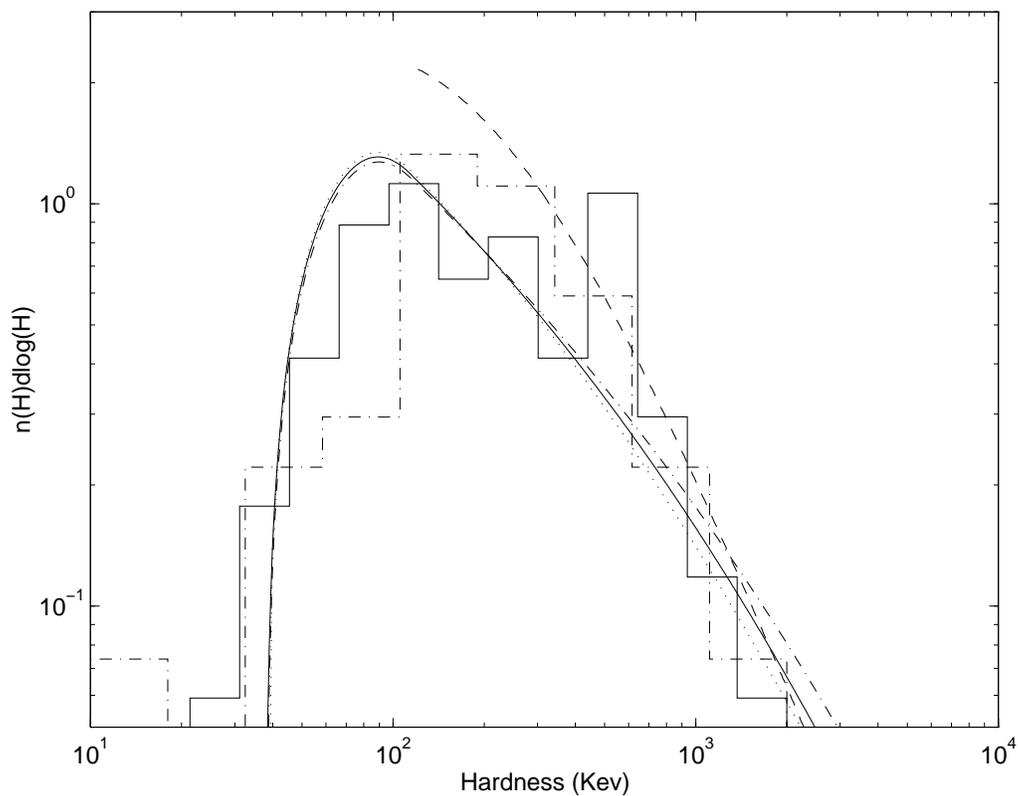}
\caption{
The observed hardness distribution for Band et. al. (1993) sample (dashed-dotted) and our sample (solid),
imposed on expected hardness distribution for intrinsic hardness distribution
with $\gamma=-0.5$,$H_{min}=120\keV$ (solid line). The dashed-dotted curve includes
effects of a DRM, and the dotted curve includes a diversity in the spectral parameter $\alpha$.
The dashed line corresponds to a hardness distribution with $\gamma=0$,$H_{min}=
120\keV$,
neglecting cosmological effects.
\label{hard_dist}
}
\end{center}
\end{figure}

The observed hardness distribution of our sample appears in
Fig. \ref{hard_dist} together with the hardness distribution of 
the sample of \cite{band}.  Fig 2. shows the distribution of the lower energy
power-law parameter $\alpha$.  We use the total duration of the bursts
to produce photon spectra.  The known hard to soft evolution causes
the hardness distribution to be softer then the hardness distribution
at the peak of the bursts, which is needed for detection
statistics. We ignore this effect. Inclusion of it will make the
intrinsic hardness distribution even harder than our estimates.

\section{The Theoretical Model}
\label{theo_hard}
To calculate a theoretical ``observed'' hardness distribution, we must
assume a model for the hardness intensity statistics of the sources.
For simplicity, we assume: (i) Standard candle in peak energy flux,
(ii) No hardness vs. intensity correlation.  We believe that
this is the simplest possible ad-hoc model.  One can easily imagine
physical processes that will lead to this situation.  For example,
within the relativistic fireball model (\cite{Piran96}) such a behavior
will arise if we keep the total energy of the fireball fixed and vary
the Lorentz factor of the relativistic motion.
(iii) We also assume  a simple form of the intrinsic
hardness distribution:
\begin{equation}
\label{H_intr}
\rho(H)d \log H = \cases{        
        0 & $H \leq H_{min}, H \geq H_{max} $ \cr
    H^{\gamma} & $H_{max}>H>H_{min}$ 
} ,
\end{equation}
where the index $\gamma$ is such that if $\gamma=0$ there are equal
number of bursts per logarithmic interval of H between $H_{min}$ and
$H_{max}$. If $\gamma>0$, then there are more hard bursts then soft
ones.  We also assume that for all bursts $\bar \alpha=-0.65$ and
$\bar \beta=-2.6$ which are the average values of our sample. (Later,
after we find the intrinsic hardness distribution which fits the
observed data the best, we check the sensitivity to a distribution of
power-law indices. See section \ref{spect_diver} ).
 
In order to produce a set of bursts with the same total energy we
calibrate the spectra by setting the constant $A$ in equation
\ref{total_energy} to hold $E_{TOT}=Const.$ fixed for all the bursts.

We calculate the distribution of observed hardness, which is
\begin{equation}
\label{N_H1}
N(H_{obs}) d \log H = \int_{H_{min}/\min(H_{obs},H_{min})-1}^{ H_{max}/H_{obs}-1}
n(z) \rho[H_{obs} (1+z) ] \Psi(z) dz ,
\end{equation}
where 
$n_z(z) = 16 \pi (c/H_0)^3 { (\sqrt{1+z} - 1)^2 (1+z)^{-7/2} } dz$ is
the proper volume of a shell extending from $z$ to $z+dz$, compensated for event count rate,
assuming constant rate of GRBs per proper time per comoving volume and $\Omega=1$.
The detection function $\Psi(z)$ states if the burst with hardness $H_{obs} (1+z)$
is observable with our detector. 
The main BATSE triggering algorithm uses only counts in the region $50\keV<E<300\keV$ 
(cf. discussion in section \ref{DRM_inc} ). 
With these assumptions,
\begin{equation}
\label{Psi}
\Psi(z;H_{obs}) = \Theta \left\{ {C_{50-300}[H_{obs} (1+z),\alpha,\beta,z] - C_{min}} \right\} , 
\end{equation}
where $C_{50-300}(H,\alpha,\beta,z)$ is the peak rate of photons the
detector receives from a source at red-shift $z$ in the interval
$50\keV - 300\keV$ (the BATSE detection window) at 1024msec. 
The $50\keV$ to $300\keV$ range (channels 2 \& 3 of BATSE) is a feature of
the BATSE triggering algorithm. Clearly a triggering algorithm 
based on different BATSE channels  
will result in different data selection effects.
The sources are
normalized as standard candles in peak luminosity using equation
\ref{total_energy}.  For simplicity we use a fixed count threshold,
$C_{min}$.  We then use the $\chi^2$ method to find which parameters
($H_{min}$,$H_{max}$,$\gamma$) fits the observed distribution the
best. 

The best fit parameter for our data set are
$H_{min}=120$\keV,$\gamma=-0.5$. The upper cut-off of the hardness
distribution, $H_{max}$ is not constrained by current data. 
This intrinsic hardness distributions agrees with \cite{NP95}, in that
the observed hardness distribution is compatible with a large number
of non-detectable MeV bursts, and the apparent upper-cut off arises
from data selection effects.

\subsection{Data selection effects.}
\label{data_selection}
\begin{figure}
\begin{center}
\plotone{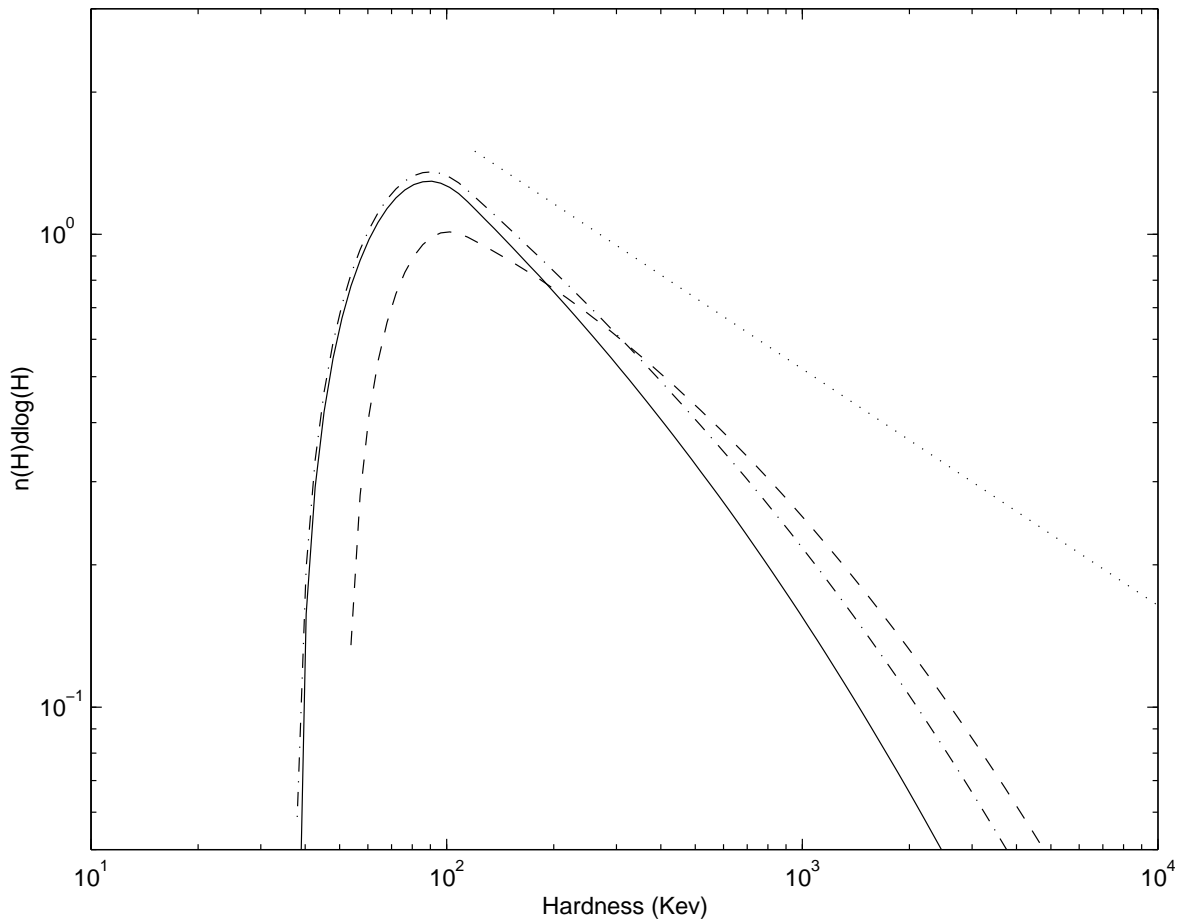}
\caption{
  The expected observed hardness distribution for different triggering
  mechanism, assuming an intrinsic  hardness distribution with
  $\gamma=-0.5$,$H_{min}=120\keV$ (un-normalized dotted line).  The
  different lines correspond to triggering on 50-300 keV (solid),
  50-2000 keV (dashed-dotted) and 300-2000 keV (dashed). 
\label{triggering}
}
\end{center}
\end{figure}

It is interesting to check whether one can overcome the data selection
effects by modifying the triggering algorithm, or are the selection
effects inherent triggering algorithm that depends on counts.  We
model three different possible algorithms, based on different photon
energy ranges: (i) 50-300 keV (currently the main BATSE algorithm)
(ii) 50-2000 keV and (iii) 300-2000 keV.  We assume that the hardness
distribution is given by Eq. \ref{H_intr} with $\gamma=-.5$ and
$H_{min}=120$ keV. We also assume that the spectrum for the underlying
  noise behaves like $\nu F_{\nu} = const$. We calculate the
  distribution of hardness expected with this intrinsic distribution
  and these three different triggering algorithms (see Fig.
  \ref{triggering}).  We find that inclusion of the the 300-2000keV
  channel increases the overall rate of observed bursts by 12\%
  compared to triggering on the 50-300 keV alone.  Using only the
  300-2000keV leads to a decrease in the total rate by 10\% (this last
  number depends rather sensitively on the lower cut off chosen in the
  intrinsic hardness distribution). What is more important is that
  even while triggering on the 300-2000 keV photons, there is still a
  large difference between the observed distribution and the intrinsic
  one (see Fig. \ref{triggering}). This shows that the inherent
  problem in detecting harder bursts is the decrease in the total
  number of photons as the hardness increases which is not compensated
  by an equivalent decrease in the noise.

\subsection{Detector Characteristics}
\label{DRM_inc}
The Detector Response Matrix translates the spectrum of incident
photons to the measured spectrum of counts. (see section
\ref{spect_estim}).  The function $C_{50-300}(H,\alpha,\beta,z)$ in
eq. \ref{Psi} ignores the DRM and uses instead the identity matrix.
In order to check this effect, we take an arbitrary
DRM (burst 3B920226) and define a modified count function
\begin{equation}
\tilde C_{50-300}(H,\alpha,\beta,z) = \sum_{i,k} DRM_{i,k} \cdot C_{\nu_i-\nu_{i+1}}(H,\alpha,\beta,z),
\end{equation}
where $\nu_i$ are the DRM photon spectra boundaries, and $k$ spans all the count channels with
energies from 50 keV to 300 keV. In table \ref{table1} we show the DRM effect on counts, for bursts with various hardnesses. 
Using this modified count function, we recalculated eq. \ref{N_H1}. 
A sample $N(H)$ distribution (see Fig. \ref{hard_dist}), with DRM inclusion, shows increasing number of 
hard bursts which results from hard photons that are measured as softer ones. 
We see that this effect does not change the distribution significantly.

\subsection{Spectral diversity}
\label{spect_diver}
\begin{figure}
\begin{center}
\plotone{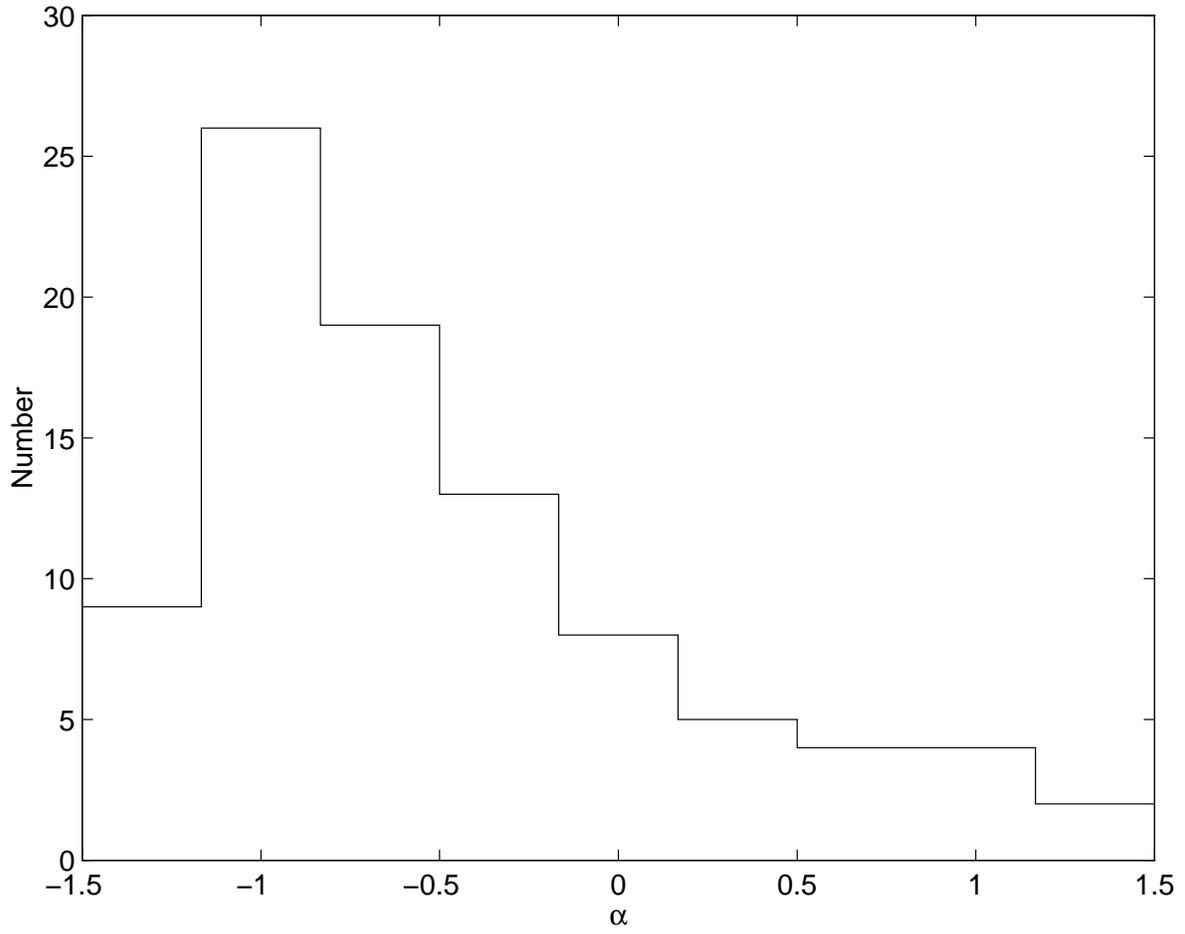}
\caption{ 
The observed distribution of the lower energy power-law parameter $\alpha$.
}
\label{alpha}
\end{center}
\end{figure}

The spectral shape of a burst in the low energy regime, i.e. the power-law parameter
$\alpha$, can determine if the detector detects the burst or not, even 
if the hardness is constant.
A hard burst with average $\alpha$ might not be detectable. A burst with the same hardness
but lower $\alpha$, has more photon in the detector window, and can be detected.
\cite{NP95} found a negative correlation between hardness and the parameter $\alpha$, 
which can be explained by this effect.
We proceed to evaluate the sensitivity of our previous calculation to diversity in $\alpha$.
We calculate the expected observed hardness distribution for intrinsic hardness distribution
and intrinsic distribution of the $\alpha$, where for the later we
take the observed one (see Fig. \ref{alpha}).
It appears from Fig. 1, that the modified hardness distribution is slightly softer, which can be
explained by the population of bursts with a higher $\alpha$ then the average one.
We find that this spectral diversity does not change our results significantly.

\section{Correlations}
\label{corr_dis}
It is generally assumed that positive correlation between fluence and hardness 
should appear if the bursts are cosmological.
However, while looking for this correlation one should beware
of correlating between parameters which have an inherent correlation induced  by their 
estimation method (\cite{Sch93}). 

For example, assuming that $\alpha$ \& $\beta$ are constants for all
bursts, equation \ref{total_energy} becomes $F \propto N \cdot H$,
where $F$ is the fluence, $N$ is the photon count, and $H$ is the
bursts' hardness. Assuming that there is no intrinsic correlation
between the photon counts and hardness, and that the distribution
function are "well behaved", we define the spread in hardness and
counts by $S_H=Var(H)/\avg{H}^2$ and $S_N=Var(N)/\avg{N}^2$
respectively. Then the correlation coefficient between fluence and
hardness is
\begin{equation}
r = { \sum_i{ ( N_i H_i - \avg{NH})  ( H_i - \avg{H} ) }  \over
    \sqrt{ \sum_i{( N_i H_i - \avg{NH} )}^2 \sum_i{( H_i - \avg{H}) }^2 } } =
\sqrt{ S_H \over S_N ( 1 + S_H ) + 1 } > 0 .   
\end{equation}
The result depends on the distribution of $H$ and $N$, but it is always
positive, and it can have an arbitrary positive value without any
intrinsic correlation.

\begin{figure}
\begin{center}
\plotone{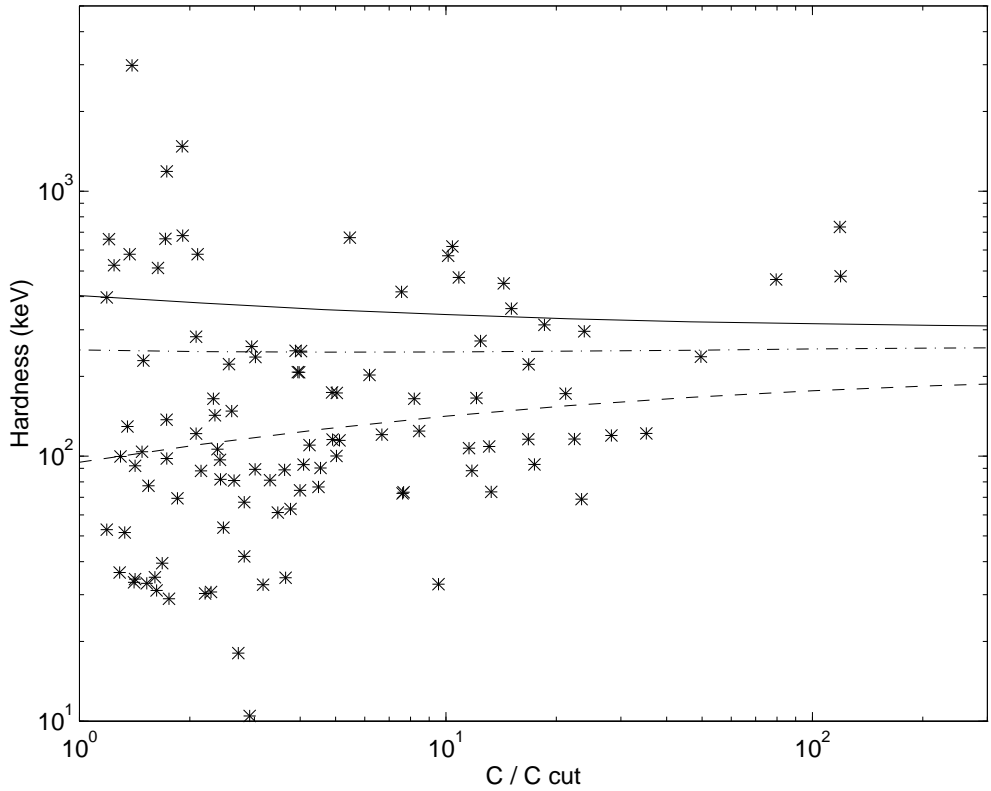}
\caption{ 
The observed hardness vs. counts distribution, with the theoretical curves for
a constant number of GRBs per decade of hardness with lower cut off only (solid), 
with lower and upper cut off (dashed) and the best fit model ($\gamma=-0.5$,$H_{min}=120\keV$) 
(dashed-dotted).
The counts are in the 1024msec channel, and $C_{cut}=289$.
\label{corr}
}
\end{center}
\end{figure}

In the case of standard candles with an intrinsic hardness
cosmological effects lead to a one to one relation between counts and
observed hardness, and a positive correlation between them is
inevitable.  However, consider a population of GRBs in a certain
red-shift $z$, with a hardness distribution.  The correlation between
hardness and counts would be {\it negative}, because ( for standard
candles ) the harder bursts emit less photons, and even less photons
inside BATSEs' triggering interval.  What should we expect from GRBs
which are spread over the universe {\it and} have an intrinsic
hardness distribution?  Fig. \ref{corr} shows the average hardness as
a function of counts for two intrinsic hardness distribution in the
form of eq. \ref{H_intr} (We prefer to use counts rather than
intensity, due to the usage of the bursts' spectra while calculating
its' intensity (\cite{Pend96})).  Both curves are for a constant
number of GRBs per logarithmic hardness interval. The descending curve
corresponds to hardness distribution with $H_{min}=100\keV$ and
$H_{max}=\infty$, and the ascending curve to hardness distribution
with $H_{min}=100\keV$ and $H_{max}=500\keV$. It is easy to see that
even a mild hardness distribution masks cosmological effects (recall
that a hardness distribution with $H_{min}=100\keV$ and
$H_{max}=500\keV$ is too narrow to fit the observed one). Thus, the
large observed hardness distribution disables the usage of
hardness-intensity relation as an independent probe for the bursts'
cosmological origin.

\begin{figure}
\begin{center}
\plotone{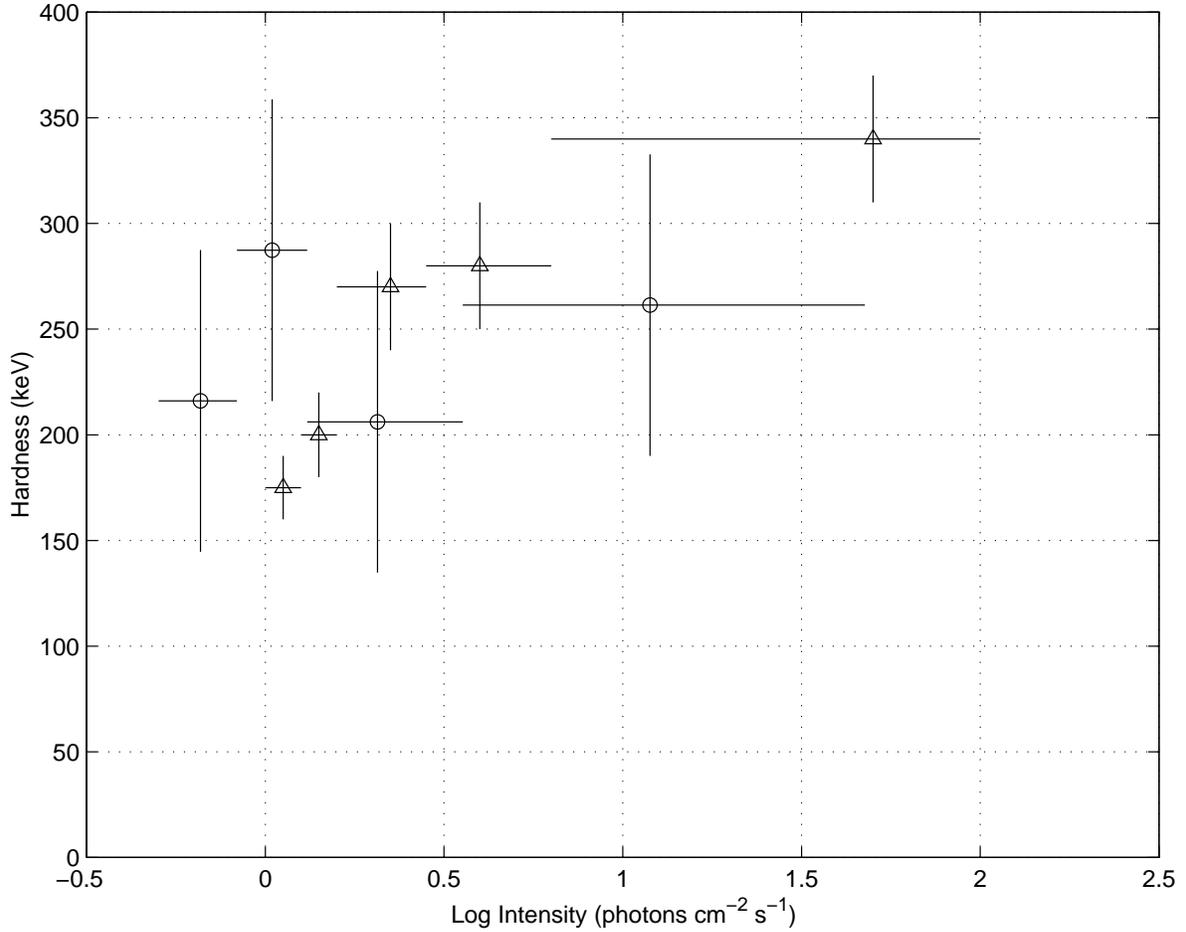}
\caption{ 
The average $\nu F_\nu$ peak energies as a function of intensity (BATSEs' peak flux) for
our results (circles) and Mallozzi \etal (1995) (triangles).
While one sample shows an increasing trend, the other one does not. Still with the large
error bars the two samples seems to be consistent.}
\label{udi_mal}
\end{center}
\end{figure}

We find a correlation coefficient smaller then $2 \times 10^{-2}$
between hardness and peak flux.  This result agrees with \cite{band}
who have found a correlation coefficient of $-8 \times 10^{-2}$.
\cite{malozzi} have found a marginally significant correlation (0.9).
In Fig. \ref{udi_mal} we compare the hardness - intensity relation in
our and \cite{malozzi} samples.  The intensity is the peak flux in
256msec channel from BATSEs' catalog.  It is not clear if the two data
sets are discrepant or not. This warrants further investigations.

\section{Discussion and Conclusions}
\label{conc}
A comparison between the expected hardness distribution for various
intrinsic hardness distributions, and the observed distribution
reveals the necessity for intrinsic paucity of soft bursts.  Any
intrinsic distribution, that does not include an intrinsic paucity in
this area, does not fit the observed distribution.  Therefore, unless
BATSE has an unexpected and unknown selection bias against soft
photons, the lower cut-off in the observed distribution is a real
phenomenon.  Using a best fit method, we found that the observed data
is best modeled by intrinsic lower cut-off at $120\keV$.

The story is, however, very different for large values of H. The data
show very small numbers of hard bursts, e.g., only two bursts out of
54 bursts in the Band et. al. (1993) sample and only five bursts out
of 136 in our sample are harder than $1$ MeV.  Nevertheless, this does
not mean that there are fewer GRBs above $1$ MeV.  The best fitted
intrinsic hardness distribution, is one with $\gamma=-0.5$, i.e. a
slowly decreasing number of bursts per logarithmic interval. Even a
model with $\gamma=0$, i.e. a constant number of bursts per
logarithmic hardness interval gives a probability of 15\% in a KS
test, which is not high but is not sufficiently low to rule out the
model.

The interpretation of the result is quite simple. There is an
observational bias against detecting bursts with $H \geq 500\keV$ by
current detectors.  Two factors operate. For bursts with a fixed
luminosity, harder bursts have fewer photons.  This makes the
detection of harder bursts difficult in any detector that is triggered
by photon counts. (If the energy of the detector noise per decade is
constant, then the the ratio between the number of photons in the
burst and in the noise remains constant.  However, the noise variance
decreases slowly with energy (square-root of the total noise), and the
signal to noise decreases.)  The decrease in sensitivity in BATSE is
even more severe since BATSE triggers on photons in the $50\keV$ to
$300\keV$ range and as the bursts become harder most of the emitted
photons are further and further away from this energy range.  A
careful comparison between the hardness distributions observed with
different triggering algorithms in BATSE 4B catalog might give some
indication on the high energy hardness distribution.

\section{Acknowledgments}
We thank D. Band for helpful discussions.
This research was supported by a US-Israel BSF grant 95-238  
and by a NASA grant NAG5-3516.

\newpage

\newpage

\begin{deluxetable}{rrrr}
\tablecaption{DRM effects - Normalized counts for bursts with various spectra.\label{table1}}
\tablehead{
\colhead{Hardness \tablenotemark{a} } &
\colhead{Counts $<$ 300keV \tablenotemark{b}} &
\colhead{Counts $>$ 300keV \tablenotemark{c}} & 
\colhead{Total counts} \\
\colhead{(keV)} &
\colhead{} &
\colhead{} & 
\colhead{} 
}
\startdata
  100.00 &        93.75   &       6.25   &     100.00  \nl
  500.00 &        35.99   &       9.23   &      45.22  \nl
  1000.00 &        17.63   &       7.13   &      24.77 \nl
  1500.00 &        11.16   &       5.50   &      16.66 \nl
  2000.00 &         7.97   &       4.41   &      12.38 \nl
\enddata
\tablenotetext{a}{Peak of $\nu F \nu$. All bursts spectrum have the same total energy with  $\bar \alpha=-0.65$ and
$\bar \beta=-2.6$,  normalized to give 100 counts for a burst with hardness of 100 keV}
\tablenotetext{b}{Counts in the 50-300 keV regime, from photons with energies $<$ 300 keV.}
\tablenotetext{c}{Counts in the 50-300 keV regime, from photons with energies $>$ 300 keV.} 
\end{deluxetable}
\end{document}